\documentclass[
reprint,
amsmath,amssymb,
aps,
superscriptaddress,
]{revtex4-2}

\usepackage{graphicx}
\usepackage{dcolumn}
\usepackage{bm}
\usepackage{commath}
\usepackage{amsmath}
\usepackage{textcomp}
\usepackage{graphicx}
\usepackage[]{wasysym}

\newcommand{\Htg}{$\mathrel{\raisebox{.3ex}{{\scriptsize $\#$}}}$}
\usepackage{hyperref}      
\usepackage{url}           
\usepackage{booktabs}       
\usepackage{amsfonts}      
\usepackage{nicefrac}      
\usepackage{microtype}      
\usepackage{lipsum}
\usepackage[euler]{textgreek}
\usepackage{color}
\usepackage[normalem]{ ulem }
\usepackage{soul}
\newcommand{\lkb}{\affiliation{Laboratoire Kastler Brossel, Sorbonne Universit\'e, \'Ecole Normale Sup\'erieure-Paris Sciences et Lettres (PSL) Research University, Centre National de la Recherche Scientifique (CNRS) UMR 8552, Coll\`ege de France, 24 rue Lhomond, 75005 Paris, France}}
\newcommand{\lpens}{\affiliation{Laboratoire de Physique de l'\'Ecole Normale Sup\'erieure, Universit\'e Paris Sciences et Lettres (PSL), Centre National de la Recherche Scientifique (CNRS), Sorbonne Universit\'e, Universit\'e Paris-Diderot, Sorbonne Paris Cit\'e, 24 rue Lhomond, 75005 Paris, France.}}

\begin{document}

\preprint{APS/123-QED}

\title{Non-invasive focusing and imaging in scattering media with a fluorescence-based transmission matrix}

\author{Antoine Boniface}
\lkb
\email{antoine.boniface@lkb.ens.fr}
\author{Jonathan Dong}
\lkb
\lpens
\author{Sylvain Gigan}
\lkb

\begin{abstract}
In biological microscopy, light scattering represents the main limitation to image at depth.
Recently, a set of wavefront shaping techniques has been developed in order to manipulate coherent light in strongly disordered materials. The Transmission Matrix approach has shown its capability to inverse the effect of scattering and efficiently focus light. In practice, the matrix is usually measured using an invasive detector or low-resolution acoustic guide stars. Here, we introduce a non-invasive and all-optical strategy based on linear fluorescence to reconstruct the transmission matrices, to and from a fluorescent object placed inside a scattering medium. It consists in demixing the incoherent patterns emitted by the object using low-rank factorizations and phase retrieval algorithms. We experimentally demonstrate the efficiency of this method through robust and selective focusing. Additionally, from the same measurements, it is possible to exploit memory effect correlations to image and reconstruct extended objects. This approach opens up a new route towards imaging in scattering media with linear or non-linear contrast mechanisms.
 
\end{abstract}

\maketitle

Propagation of light in materials with refractive index inhomogeneities, such as biological tissues, results in scattering. In such disordered media, ballistic light exponentially decreases with penetration depth, which limits the scope of conventional optical microscopy. At depth, coherent light is affected by multiple scattering and produces very complicated interference patterns, known as speckle \cite{goodman1976some}. Recent years have witnessed many advances in the ability to coherently manipulate this figure of interferences owing to the availability of spatial light modulators (SLMs) \cite{mosk2012controlling,rotter2017light}. In particular, they enable refocusing light to a diffraction-limited spot \cite{vellekoop2007focusing}. These techniques often rely on optimizing the incident wavefront such that it maximizes a feedback signal emitted from the target focus point. A key constraint for biological imaging at depth, is that the measurement of the feedback has to be non-invasive. Several strategies using acoustics or non-linear fluorescent guidestars have been proposed for this purpose \cite{horstmeyer2015guidestar}. However, most of them are only able to form a single focus at a given output position \cite{lai2015photoacoustically,Boniface:19,katz2014noninvasive,yaqoob2008optical}, which limits the acquisition speed or the field-of-view. 

Deterministic focusing of light on multiple targets is optimally achieved with a transmission matrix (TM) that linearly relates the input field to the output field \cite{popoff2010measuring}. However, its non-invasive measurement remains very challenging. Indeed, even if each target has its own optical response, what is measured in epi-detection is the back-scattered emission, thus spatially and temporally mixed. To overcome this limitation, optics has been combined with acoustics to coarsely locate each target \cite{chaigne2014controlling,katz2019controlling}, which enables the reconstruction of a TM but requires complicated acousto-optical setups. Another powerful approach relies on the measurement of a time-gated matrix in reflection \cite{jeong2018focusing, badon2016smart}, but is based on retro-reflected ballistic photons, hence limited in depth.

Linear fluorescence remains an essential technique in microscopy because systems are fairly inexpensive and easy to handle. As such, it remains a staple tool in biology and biomedical sciences. It has enabled imaging of cells and sub-microscopic cellular components with high spatial resolution, specificity, contrast and speed. Combined with light-sheet or structured illumination microscopy, linear fluorescence allows sectioning and imaging at moderate depth \cite{power2017guide,bouchard2015swept}. Although imaging fluorescent objects through thin scattering media can be done thanks to the memory effect \cite{katz2014non, hofer2018wide}, a general method to focus and image a fluorescent object at depth is still missing. 
\begin{figure*}[!t]
\centering
\includegraphics[width = 420px] {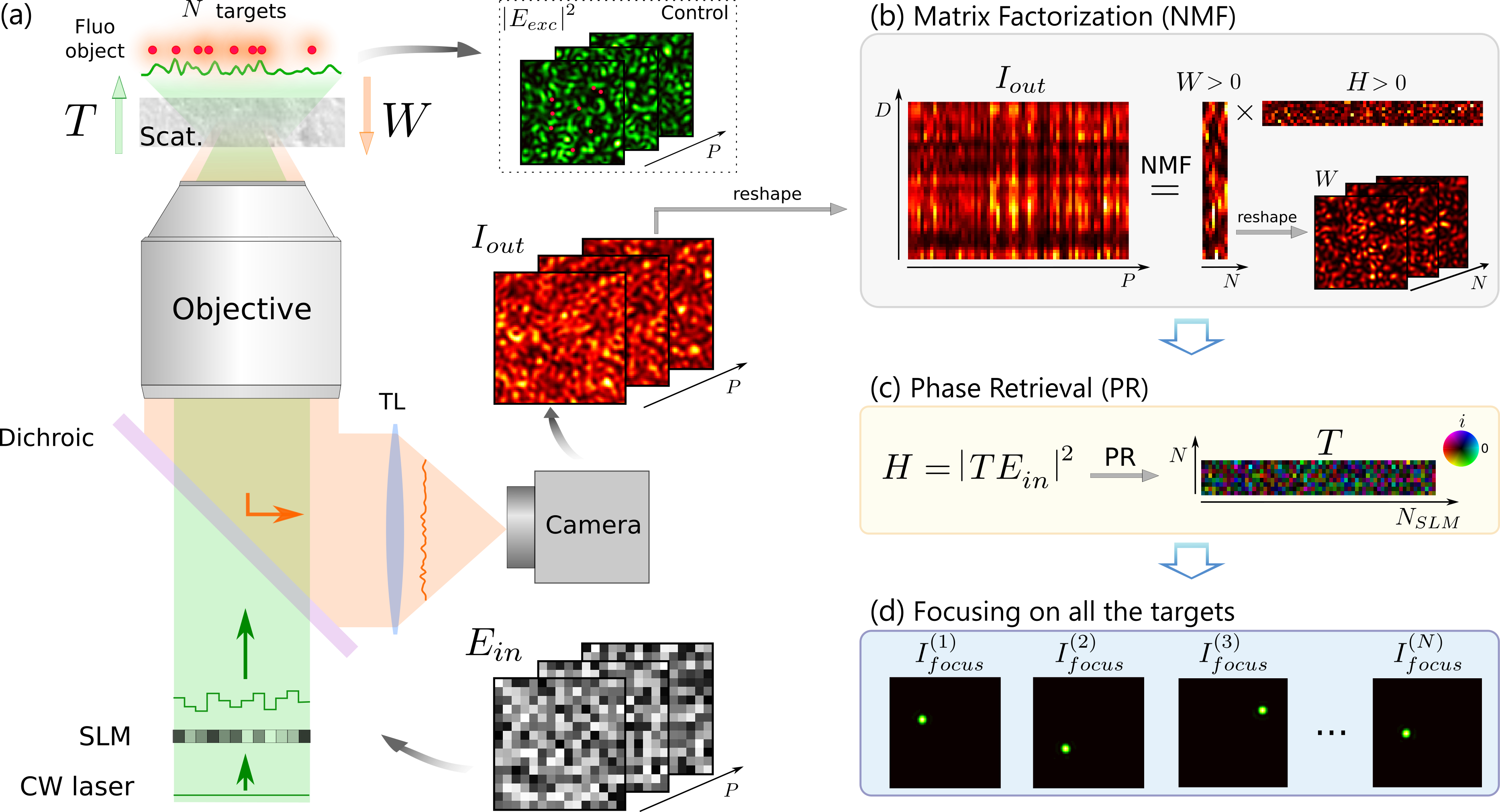}
    \caption{Double-TM reconstruction principle - simulation results. (a) Schematic view of the experimental setup. Coherent light is sent on a fluorescent object, made of $N$ targets, hidden behind a scattering medium. A speckle field, $E_{exc}=TE_{in}$, illuminates the object which emits a fluorescence signal in return. A portion of the latter is back-scattered by the medium and epi-detected on a camera, $I_{out}=W\abs{E_{exc}}^2$. TL: tube lens, Scat.: scattering medium. (b) $I_{out}$ is a sequence of $P$ fluorescent speckles recorded for different inputs, $E_{in}$. This matrix admits a rank-$N$ factorization $I_{out}=WH$, where $W$ and $H$ are unknown positive matrices. NMF is used to retrieve them. $W$ is an intensity-TM describing the fluorescence propagation from the object to the camera. (c) $H$ describes light propagation from the SLM to the object and can be written as $H=\abs{TE_{in}}^2$, where $T$ is a field-TM. An additional step of phase retrieval gives access to $T$. (d) Phase conjugation of $T$ is used to selectively and non-invasively focus light on all the targets of the fluorescent object. This focusing ability is used to quantify the quality of the double-TM reconstruction.  
}
\label{principle}
\end{figure*}

Here, we report on a robust TM approach for fluorescence imaging through a relatively strong scattering medium, in a non-invasive way. The technique relies on shining a sequence of known wavefronts on a fluorescent object hidden behind a scattering medium, and collect in reflection the corresponding low-coherence fluorescent speckles back-scattered by the medium. From this set of input-output information, we are able to computationally retrieve both \textit{(i)} the ingoing field-TM for the excitation light, and \textit{(ii)} the outgoing intensity-TM for the fluorescence light. We demonstrate robust and selective focusing across all the object, both on beads and on more complex fluorescent objects. Finally, if the medium exhibits limited memory effect, we show that an image of the object can be retrieved, even when its size exceeds the memory effect range.

\vspace{7mm}

The experimental apparatus is depicted in FIG. \ref{principle}a. A coherent beam of light is first modulated in phase by an SLM and directed through a scattering medium onto a fluorescent object made of several emitters. To describe both the ingoing and outgoing light propagation, we use a transmission matrix formalism. A field-TM, denoted $T$, connects the input field $E_{in}$ (specifically the phase pattern displayed onto the SLM) to the field at the position of the N targets. Thus, the speckle intensity in the plane of the fluorescent object reads $\abs{E_{exc}}^2=\abs{TE_{in}}^2$. Once excited, each target fluoresces proportionally to its illumination. This low-coherence signal is back-scattered by the medium and can be non-invasively measured with a camera placed in reflection. It can be written as $I_{out} = W \abs{E_{exc}}^2$, where $W$ is an intensity-TM, linking the $N$ targets to the $D$ pixels of the camera via their respective fluorescent eigen-patterns. We define here \emph{eigen-patterns}, as all the independent speckles, each single target generates on the camera. 
It is worth stressing that the measurement is made in reflection only, thus entirely non-invasive. A control camera placed on the far side of the sample allows to monitor the excitation patterns $\abs{E_{exc}}^2$ at the object plane.

Our technique relies on exciting the sample consisting of the scattering medium \emph{and} the fluorescent object with a variety of $p = 1, \ldots,P$ random input phase patterns $E_{in}(p)$ and collecting the fluorescence responses $I_{out}(p)$ reflected on the same side. 
For all $p = 1, \ldots,P$, $I_{out}(p)$ can be written as:
\begin{equation*}
    I_{out}(p) = W \abs{E_{exc}(p)}^2 = W \abs{TE_{in}(p)}^2
\end{equation*}
$I_{out}(p)$ corresponds to a low contrast speckle because, first, the fluorescence emission is broadband and second, the $N$ beads generate $N$ different speckles that partially average out \cite{goodman1976some}. Nevertheless, the decrease in contrast due to the number of beads is relatively slow and scales as $\sqrt{2/N}$ in the case of linear fluorescence (see Supplementary \ref{sub2p_cont}).
The overall output $I_{out} \in \mathbb{R}_{+}^{D \times P}$ can be written as a rank-$N$ product of two positive matrices $I_{out}=WH$ (with $N \ll D, P$), where both $H = \abs{TE_{in}}^2$ and $W$ are real positive matrices. This corresponds exactly to the framework of Non-negative Matrix Factorization (NMF) that we therefore use to estimate $W$ and $H$ from $I_{out}$, see FIG. \ref{principle}b. Thanks to its robustness and interpretability, this framework has already been applied in many settings \cite{moretti2019readout}, including the fluorescent readout of neuronal activity \cite{berry2007algorithms, zhou2018efficient}, but has never been associated with wavefront shaping yet. In a second step, a phase retrieval (PR) algorithm allows retrieving the ingoing field-TM, $T$ \cite{Dremeau:15, metzler2017coherent}, from $H$. 

\begin{figure}[!t]
\centering
\includegraphics[width=\linewidth]{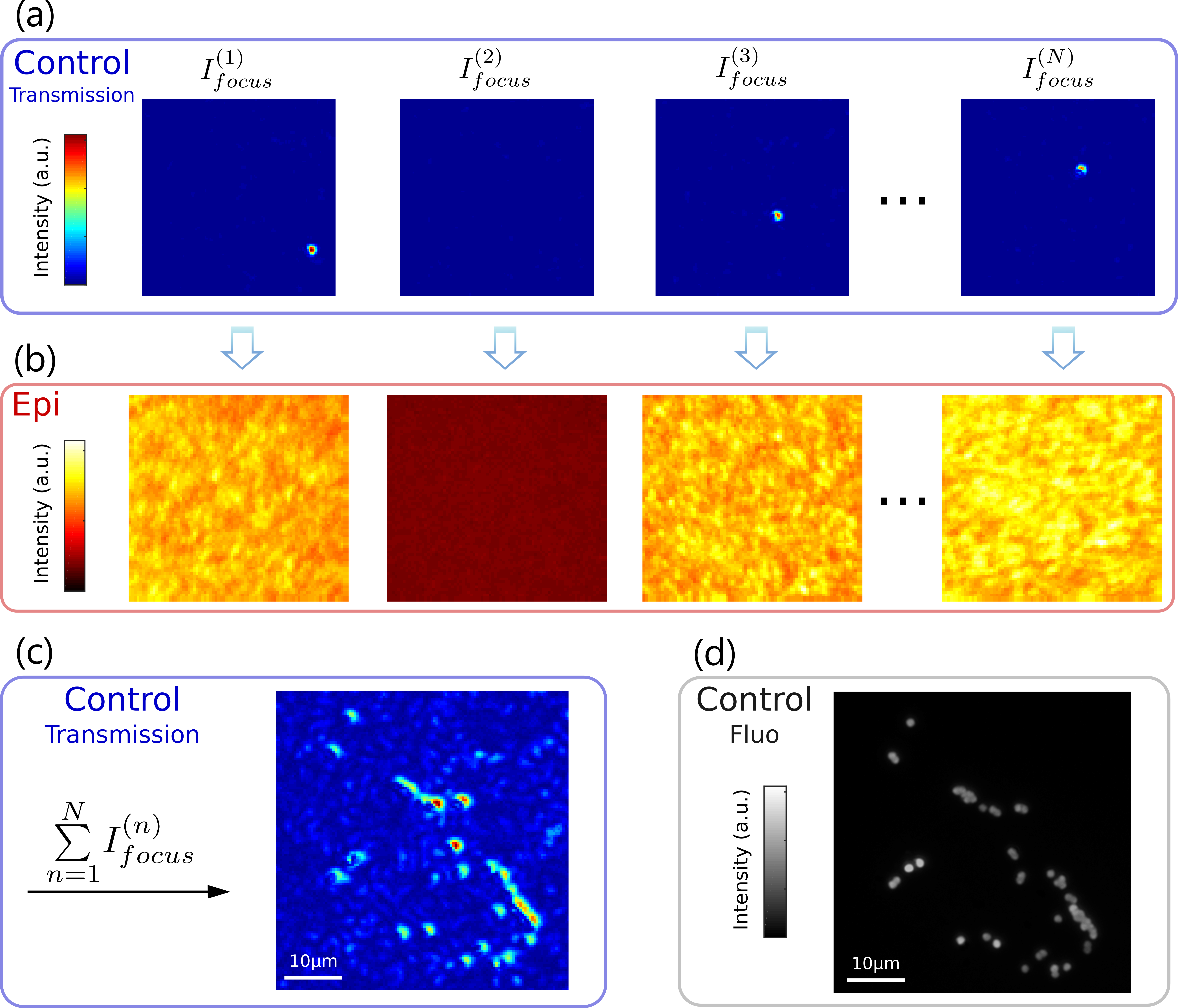}
    \caption{Experimental Results: TM reconstruction and focusing on beads through a ground glass diffuser. From the experimentally measured $I_{out}$, the field-TM $T$ is reconstructed using NMF and a phase retrieval algorithm. To prove the success of the reconstruction, we investigate in the ability of $T$ to focus excitation light. A rank $r=30$ is set as input parameter for the NMF. (a) Examples of focus spots obtained after phase conjugation of $T$. Since the rank is overestimated ($r>N$), the NMF may generate spurious eigen-patterns and reconstruction thus partially fails. Corresponding points do not generate a focus, like $I_{focus}^{(2)}$. (b) This validation step can be performed non-invasively by looking at the spatial variance of the epi-detected fluorescent patterns. (c) Sum of all the $r=30$ intensity images recorded on the control camera. It shows that our technique is able to focus on most targets. (d) Fluorescence image of the object obtained without the scattering medium for control only. Here $P=15360$ -- Acq. speed$ = 50Hz$ -- $N_{SLM}=1024$. 
}
\label{results}
\end{figure}
\vspace{5mm}
Experimentally, we first performed the measurement with fluorescent objects made of $1$\textmu m beads, placed behind holographic diffusers. Once a series of input patterns have been displayed and the corresponding fluorescence images recorded, the matrix $I_{out}$ is factorized into two low-rank matrices thanks to NMF. The rank $r$ of the factor matrices is the main input parameter required to run the algorithm. In principle, $r$ should correspond to the number of independent targets $N$ in the system, which is unknown in such reflection configuration. Nevertheless as discussed in Methods, an upper bound for $r$ can be easily estimated from $I_{out}$, and is sufficient to identify all the $N$ targets. An additional step of phase retrieval estimates the field-TM $T$ that links the SLM pattern to the plane of the beads.

Focusing light using phase conjugation provides a method to check the quality of the NMF + PR pipeline. On FIG. \ref{results}, this focusing capability is shown with images monitored on the control camera, but also non-invasively from the observation of back-scattered fluorescence.
When light is successfully focused on a target, the spatial variance of the fluorescence speckle increases \cite{Boniface:19}. Since we typically overestimate the rank $r>N$, the NMF may generate spurious eigen-patterns (which do not focus the illumination) and duplicates. Looking at the spatial variance and the correlation between fluorescent patterns allows to identify both (see Supplementary \ref{supp_varCor}). We thus validate the ability to accurately reconstruct the ingoing and outgoing transmission matrices and deterministically focus on every beads.

\begin{figure}[!b]
\centering
\includegraphics[width=\linewidth]{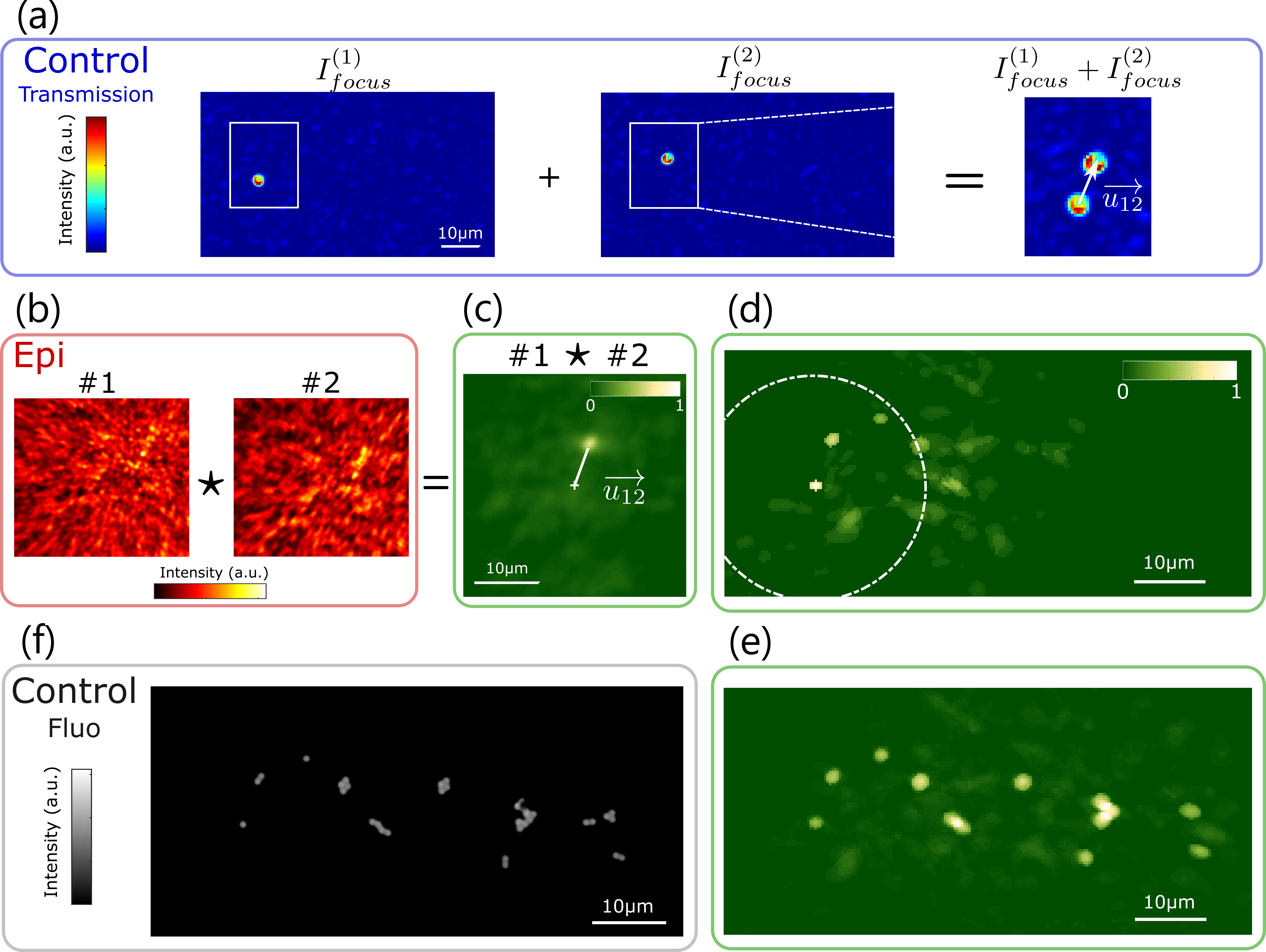}
    \caption{Imaging through scattering media using back-scattered fluorescence and finite memory effect. (a) Images recorded on the control camera when the illumination is focused on two different beads of an extended object represented in (f), where the scattering medium is made by two surface diffusers separated by $0.78$mm. We define $\protect\overrightarrow{u_{12}}$ as the relative displacement between the two foci (i.e. beads). (b) If the two beads are within the same ME patch their two fluorescent patterns, \Htg1 and \Htg2, are spatially shifted by $\protect\overrightarrow{u_{12}}$. (c) $\protect\overrightarrow{u_{12}}$ is estimated from cross-correlation \Htg1$\star$\Htg2. (d) This operation is repeated between \Htg1 all the other eigen-patterns \Htg $j$ with $j = 1, \ldots,13$. If bead \Htg $j$ is within the ME patch of \Htg1, indicated by the dashed circle, a peak of correlation appears and a sub-part of the object is retrieved. (e) To obtain the full extended object, the reconstruction is done as in (d) for all the ME patches. The result is in good agreement with the ground truth (f) which is a fluorescence image recorded without scattering medium. $P=14336$ -- Acq. speed $=20Hz$ -- $N_{SLM}=1024$.}
\label{imaging}
\end{figure}

\vspace{5mm}
The double-TM reconstruction can be applied in principle whatever the depth and scattering properties of the medium, as long as it provides a measurable fluorescent speckle. In the following, we show that, if there is some memory effect (ME), the technique allows not only focusing but also fluorescence imaging at depth by looking at the correlations between fluorescent eigen-patterns.

In essence, two beads within the ME should exhibit translated fluorescent patterns with a shift equal to their relative distance. By cross-correlating the fluorescence patterns, which are recorded in epi while displaying the $r$ focusing patterns onto the SLM, it is possible to retrieve a distance map between all the beads. In FIG. \ref{imaging}, we show an example of such reconstruction with an object much larger than the ME range (see Supplementary \ref{subME}). Interestingly, computing successively these pairwise correlations between close targets allows retrieving the full object, well beyond a single ME patch. All the beads are thus reconstructed as long as their ME patches have some overlap. In FIG. \ref{imaging} we show reconstruction of an object extending approximately three times the ME range. Note that using directly the $W$ patterns from the NMF works but did not provide here as good results (see Supplementary \ref{supp_dist}).

\begin{figure}[!t]
\centering
\includegraphics[width=\linewidth]{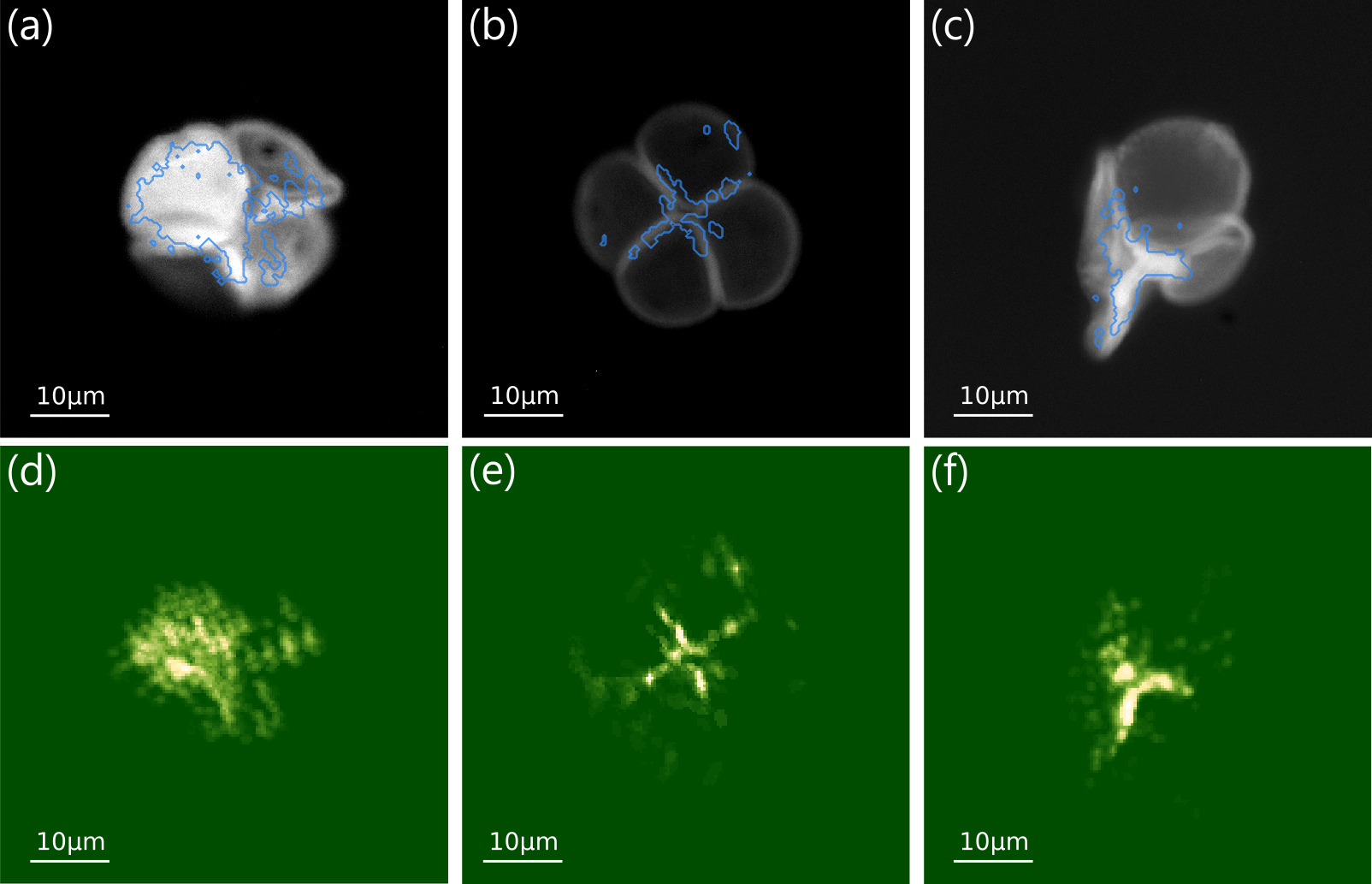}
    \caption{Reconstruction examples for continuous and volumetric fluorescent objects through a ground glass diffuser. (a)-(c) Fluorescence images of different pollen seed structures taken after removing the diffuser. The blue line is a contour plot (at $10\%$ of the maximum intensity) of the reconstructed object. (d)-(f) Reconstruction of the object with NMF + PR approach. The rank for the NMF is overestimated at $r=80$ for (d) and $r=150$ for (e) and (f). Note that only the high variance eigen-patterns, are cross-correlated. In the three cases, $P=5120$ fluorescent images are recorded at 50Hz, $N_{SLM}=256$.}
\label{pollen}
\end{figure}

Finally, to demonstrate that our technique can also be used with 3D continuous objects, we tested it on biological objects, here fluorescence-stained pollen grains. The whole process, from the acquisition to the reconstruction, is similar to what is presented in FIG. \ref{imaging}.  FIG. \ref{pollen}a-c shows fluorescent images of three different pollen grains taken without the diffuser. The blue lines are contours of the reconstructed image at $10\%$ of the maximum intensity, showing that the high-intensity features of the object are faithfully retrieved. The full reconstructed images are presented in FIG. \ref{pollen}d-e.

\vspace{7mm}

Several conditions should be met in order to successfully operate our non-invasive technique. A first point is the required number of patterns to accurately retrieve the double-TM.
As detailed in Supplementary \ref{sub1}, reconstructing $W$ can be done even with a low number of patterns related to the complexity of the object (number of separate emitters). On the other hand, $T$ is recovered through an additional step of phase retrieval which requires a larger number of patterns, related to the number of SLM pixels. For example, experiments of FIG. \ref{results} and \ref{imaging} require $P \simeq 15000$ patterns, which at $50 Hz$ (limited by the exposure time of the camera) corresponds to $\sim 5$ minutes.
Diminishing the number of SLM pixels as in FIG. \ref{pollen} where $N_{SLM}=256$ reduces the acquisition time to few tens of seconds, which should be compatible with stability time of ex-vivo biological tissues \cite{jang2015relation}. 

Another important aspect is the complexity of the object that can be reconstructed. Here we demonstrate focusing and imaging on multiple beads, but also on continuous and even 3D objects. One limitation is the contrast of the measured speckle, that decreases with the complexity (number of separate emitters) of the object, but only with a mild squareroot dependency. Here, we only reconstruct the 2D shape of the object, but in principle the technique could be extended to 3D imaging \cite{okamoto2019noninvasive}.  
In tissues, a general problem is the background fluorescence, that could be tackled via appropriate sparse staining or acoustic tagging of a small region \cite{katz2019acoustoptic,wang2012deep}.

Regarding imaging, our technique is limited by the spatial sparsity of the object rather than its size that can be much larger than the memory effect range. We propose in the Supplementary \ref{subMDS}, an alternative algorithm based on Multi-Dimensional Scaling (MDS) that offers some advantage in terms of noise robustness. But still, the major limitation is that isolated targets (without correlations with others) cannot be correctly located.  

We focused here on linear fluorescence contrast, but the technique should readily generalize to any incoherent linear mechanisms, such as spontaneous Raman. Non-linear incoherent processes should also be possible (as shown in Supplementary \ref{sub2p_cont} for 2-photon fluorescence), which should benefit from a higher contrast and lower background, at the cost of a lower overall signal. 

In conclusion, we have presented a completely non-invasive computational strategy to characterize light propagation in and out of a scattering medium based on linear fluorescence feedback only. It allows both focusing at depth and, providing some memory effect is present, imaging of an extended object.
The method is very simple, robust, and provides a promising route towards deep fluorescence imaging beyond the ballistic regime. It should be applicable to a large variety of contrast mechanisms.

\section*{ACKNOWLEDGMENTS}
The authors thank Claudio Moretti for fruitful discussions and constructive comments.

\section*{ADDITIONAL INFORMATION}
This research has been funded by the European Research Council ERC Consolidator Grant (Grant SMARTIES - 724473). S.G. is a member of the
Institut Universitaire de France.

\cleardoublepage
\section*{METHODS}
\subsection*{Experimental setup}
\hspace{5mm} 
A continuous-wave laser ($\lambda$=532nm, Coherent Sapphire) is expanded on a phase-only MEMS SLM (Kilo-DM segmented, Boston Micromachines), such that all the $N_{SLM}=1024$ segments can be used. Once modulated, the beam is directed through the illumination objective (Zeiss W "Plan-Apochromat" $20\times$, NA 1.0 ) to excite the fluorescent object made of orange beads (540/560nm, Invitrogen FluoSpheres, size $1.0$\textmu m) or pollen seeds (Carolina, Mixed Pollen Grains Slide, w.m.) placed on top of the scattering medium. The excitation beam (diameter $< 6$mm) underfills the objective back aperture (diameter $20$mm) which reduces the actual illumination NA. It results that the speckle grain size at the fluorescent object plane is around $1$\textmu m. The SLM is imaged to the back focal plane of the microscope objective. The scattering medium is not the same in all the experiments in order to control the memory effect. In the experiment presented in FIG. \ref{results} we use a ground glass (Thorlabs, DG10), in FIG. \ref{imaging} we use two holographic diffusers (Newport 1° + Newport 10°) and in FIG. \ref{pollen} only one holographic diffuser (Newport 1°).

Part of the 1-photon fluorescence emission is back-scattered by the medium and epi-detected on a first camera: CAM1 (sCMOS, Hammamatsu ORCA Flash). Recording of the $P$ fluorescence images is the slowest step throughout the acquisition process; it is between 20 and 50 Hz depending on the scattering medium and the fluorescent sample. Once acquired, raw images are cropped (such that one image contains roughly few tens of speckle grains). Then a high pass Gaussian filter removes the background which significantly improves the contrast. The corresponding data form a matrix $I_{out}$ which is later processed with the algorithm to reconstruct the two TMs. 
We use a dichroic mirror shortpass 550nm (Thorlabs) and two other filters (F): a 532nm longpass (Semrock) and a 533nm notch (Thorlabs). A second microscope objective (Olympus "MPlan N" $20\times$, NA 0.4), placed in transmission, provides an image of the plane of the beads, onto a CCD camera CAM2 (Allied Vision, Manta). This part of the setup is for passive control only. It allows us to correctly position the beads using a white light source (Moritex, MHAB 150W), but also to monitor illumination speckles $\abs{E_{exc}}^2$. 

In the first experiment presented in FIG. \ref{results}, $P=15360$ different random inputs are generated and corresponding fluorescence images of size $D=50\times52=2600$ pixels were recorded. In the second experiment presented in FIG. \ref{imaging}, $P=14336$ different random inputs were generated and corresponding fluorescence images of size $D=70\times64=4480$ pixels were recorded on the camera in epi. 
In the third experiment presented in FIG. \ref{pollen}, $P=5120$ different random inputs were generated and corresponding fluorescence images were recorded on the camera in epi.  

The experimental setup is shown in Supplementary \ref{sub_exp}.

\subsection*{NMF + PR algorithm}
\hspace{5mm} 
Before factorizing $I_{out}$ with the NMF algorithm, the rank $r$ of the low-rank factor matrices needs to be determined. The latter is related to the number of fluorescent beads in the sample and is not known in our reflection configuration. We estimate it by looking at the residual error $\norm{I^{fluo}-WH}_F$ as a function of the rank $r$. Its plot should have a typical change of slope, as described in \cite{hutchins2008position}. It provides a good estimate for the rank of $I_{out}$.
However, when the number of targets $N \gtrapprox 10$ we experimentally observe that the change of slope cannot be determined with good accuracy. As detailed in Supplementary \ref{sub_reconstruction}, we decided to take the upper bound and remove spurious values afterwards. 

For the NMF, we use the $nnmf$ Matlab function with default parameters. In particular, matrices for the initialization are random.

For the PR we use an algorithm very similar to \cite{candes2015phase}, involving a spectral method to obtain a good initial estimate for the subsequent gradient descent iterations, except that we use a refined spectral initialization to speed up the convergence \cite{mondelli2019fundamental, luo2019optimal}.

Simulation codes are available at: \url{https://github.com/laboGigan/NMF_PR}

\cleardoublepage
\section*{Supplementary Information}
\subsection{Experimental setup}
\label{sub_exp}
\begin{figure}[htbp]
\centering
\includegraphics[width=150px]{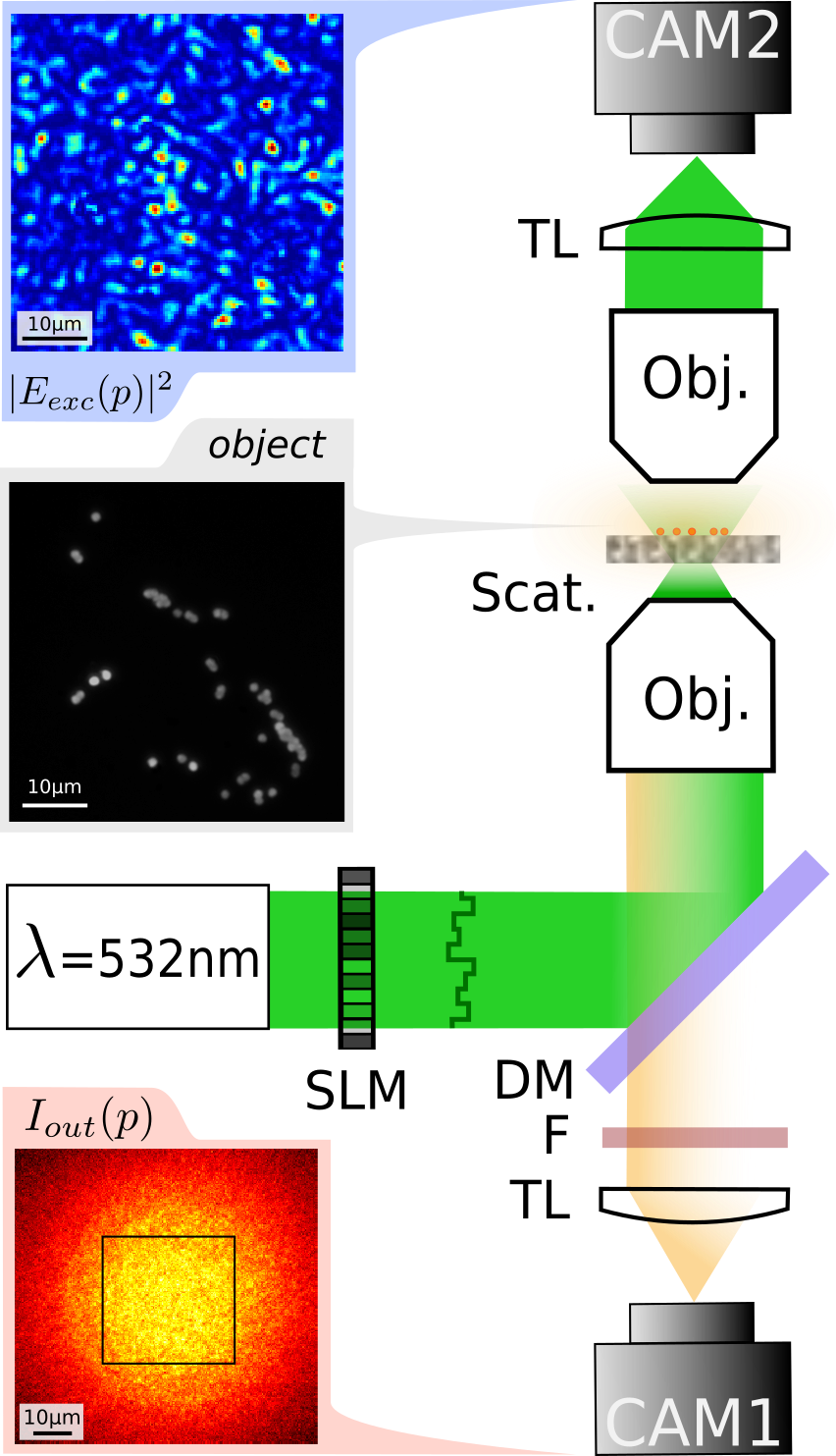}
    \caption{Scheme of the experimental setup -- From top to bottom -- $\abs{E_{exc}(p)}^2$ is the illumination speckle in the plane of the fluorescent object recorded on the control camera CAM2. \textit{object} is a fluorescence image of the object obtained on CAM1 without the scattering medium. $I_{out}(p)$ is a typical fluorescent speckle image epi-detected on CAM1. The dark square indicates the typical the cropped region we use as input data for the algorithm. DM: dichroic mirror, TL: tube lens, F: filter, Scat.: scattering medium. 
}
\label{setup}
\end{figure}
\subsection{TM-reconstruction robustness}
\label{sub_reconstruction}
\textit{In this subsection, the experimental data corresponds to the ones presented in the FIG. \ref{results} of the manuscript.}
\vspace{7mm}

\begin{figure}[t!]
\centering
\includegraphics[width=\linewidth]{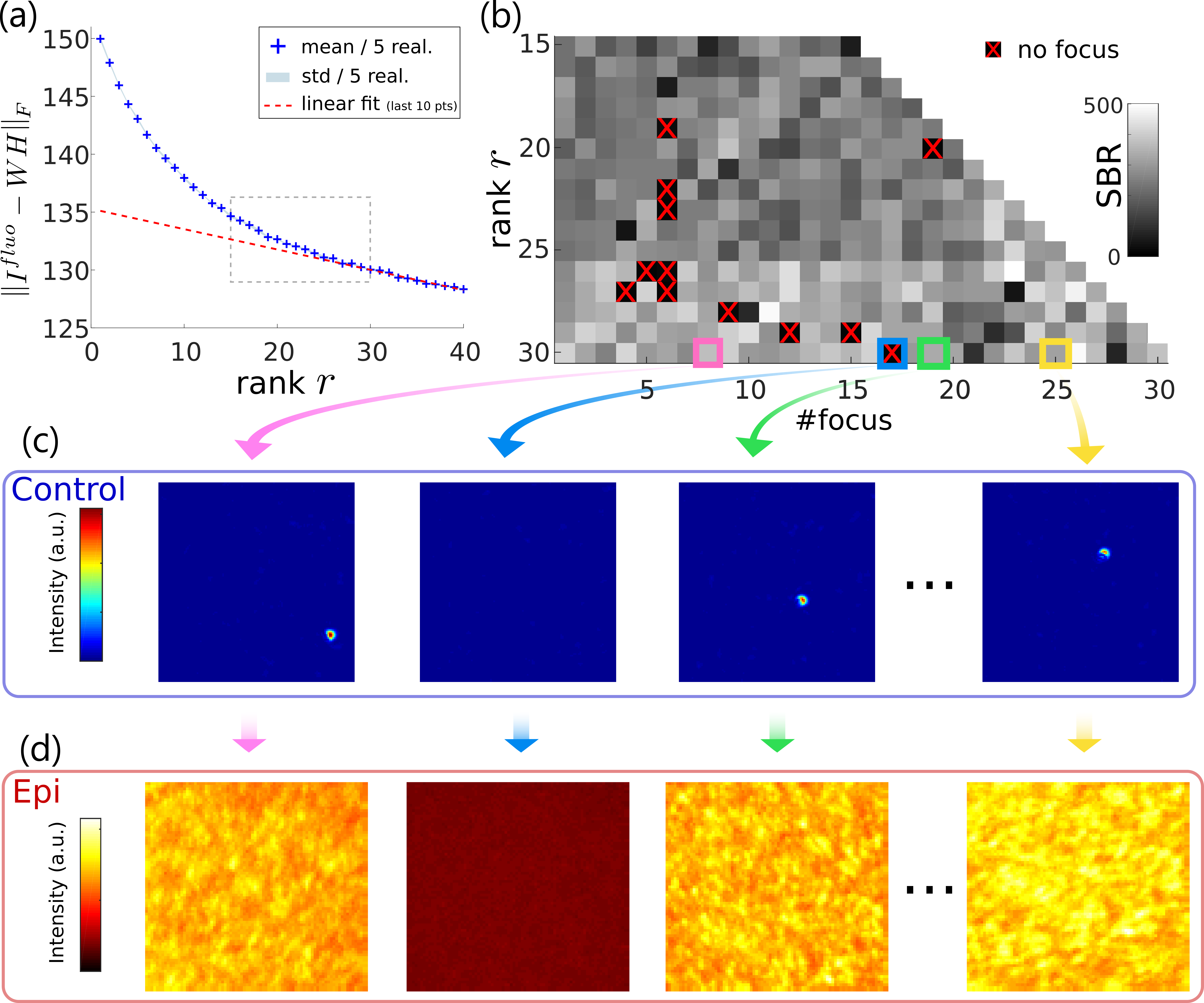}
    \caption{Experimental Results. (a) Frobenius norm of the NMF residual for rank estimation. When the latter is minimized the rank of $I_{out}$ is approximately found. It should correspond to $N$, the number of targets. Here, we have only a rough estimation of the rank of $I_{out}$. (b) TM reconstruction for different rank $r \in [15;30]$. Whatever the rank $r$ set for the reconstruction all the 16 TM are able to focus light. The only difference is the number of different spatial positions they can achieve. (c) Examples of foci obtained after phase conjugation of $T$ for $r=30$. Over the 30 lines of $T$, only one \Htg17, is not well reconstructed since it generates a speckle instead of a focus. (d) These spurious data can be tracked, non-invasively, by looking at the corresponding epi-detected fluorescence pattern. Its spatial variance is much lower than the one when the bead is successfully focused.
}
\label{supp_2910_1109}
\end{figure}

\hspace{5mm} 
To prove the robustness of our method, we decided to reconstruct $T$ for different ranks $r \in [15;30]$. 
We thus run the NMF 16 times which provides 16 pairs of matrices $\{W,H\}$. From $H = \abs{T E_{in}}^2$ and knowing SLM patterns $E_{in}$ we can recover $T$ with phase retrieval.
Repeating this procedure for all the 16 matrices $H$ gives rise to 16 different field-TMs of size $r \times N_{SLM}$, that can be used to focus light.
As a result, all the TMs whatever their rank $r$, are able to focus light, FIG. \ref{supp_2910_1109}b. This proves that the algorithm is robust and does not produce outliers if the rank is not exact. There are only 12 points with no focus at all and almost 94$\%$ of the focus spots have an SBR higher than 100. 
Looking at the variance of the fluorescence patterns on CAM1 enables us to track the points where reconstruction fails using the approach developed in \cite{Boniface:19}. As done in the manuscript, our strategy consisted of overestimating the rank $r>N$ and removing spurious data afterwards.

\subsection{Algorithm Convergence}
\textit{In the following subsection, an additional simple experiment with only $N=4$ targets is conducted, in order to study in more details algorithms input requirements and output results.}
\vspace{7mm}

\label{sub1}
\hspace{5mm}
In the following subsection, we aim at experimentally investigating the different parameters set for the algorithm and how they affect the TM reconstruction. 
\begin{figure}[htbp]
\centering
\includegraphics[width=\linewidth] {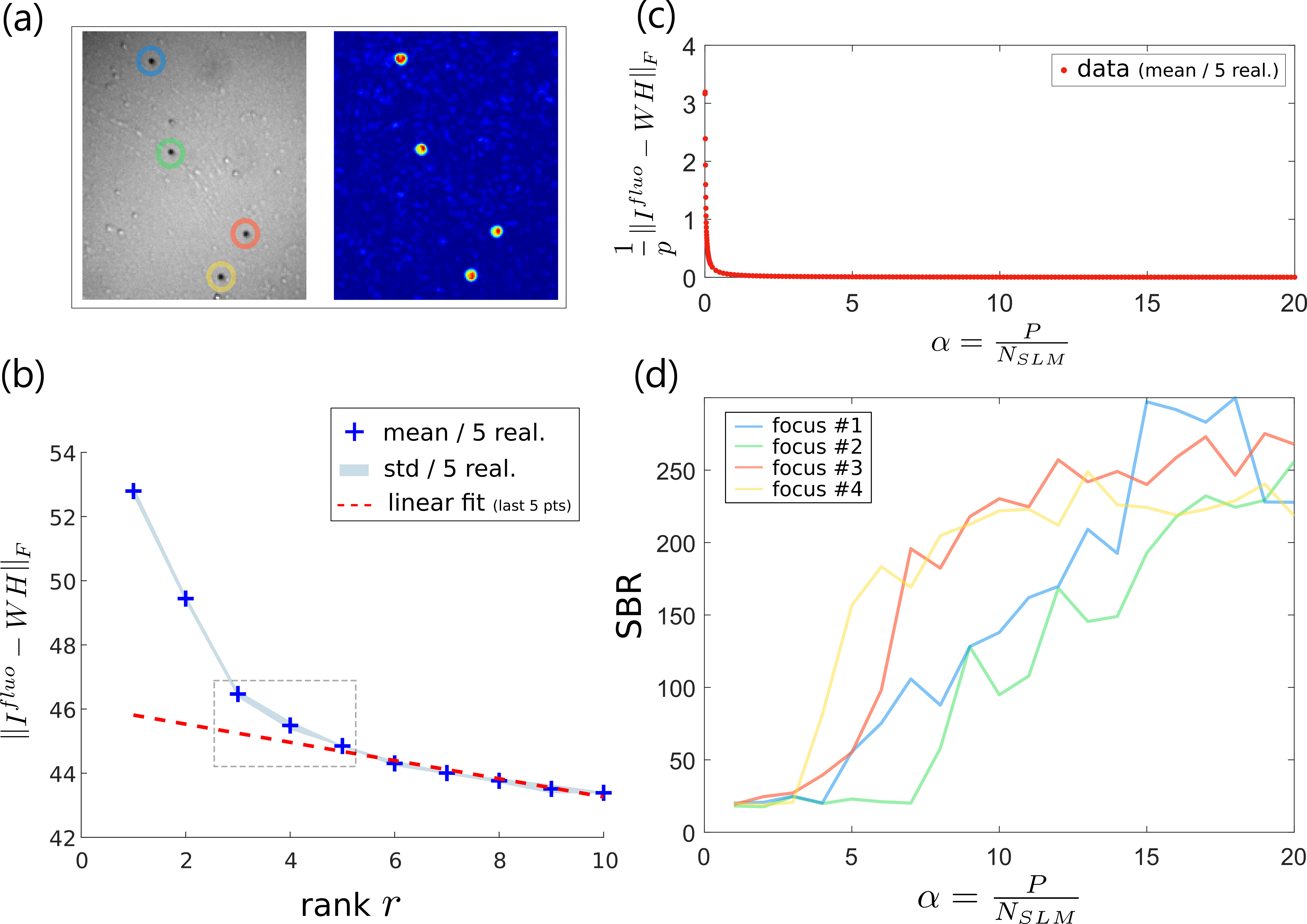}
    \caption{(a) Study of an experimental sample made of 4 targets. Once the TM is reconstructed with our computational pipeline NMF + PR, light is successively focused on all the targets. Only the sum of all the foci is represented. (b) Frobenius norm of the NMF residual for rank estimation. (c) NMF error with respect to $P$ the number of input data. (d) SBR of the generated focus after phase conjugation of $T$ as a function of input data used for the phase retrieval step.}
\label{ranksupp}
\end{figure}
In this particular study, the sample is only made of 4 beads which speeds up most of the computational steps. At first, one needs to provide the rank $r$ to factorize $I_{out}$ into $W$ and $H$. As already mentioned in Methods, this parameter can be estimated from the residual error $\norm{I_{out}-WH}_F$ of the NMF itself. A change in the slope is noticed around $r=4$ which is in good agreement with $N$. Knowing $r$, several NMF are done for different number of input patterns $p$. It shows that only few patterns ($<100$) are required to significantly reduce the error: the $r=4$ eigenvectors of $I_{out}$ are found. The last computational step consists in retrieving the phase from the NMF intensity results to get the transmission matrix. This phase retrieval problem is much more demanding in terms of number of patterns required. To highlight this point, we thus have processed the reconstruction of several TM with a different oversampling ratio $\alpha = P /N_{SLM}$, as this parameter defines the performance of recent phase retrieval algorithms in the large-dimensional limit \cite{metzler2017coherent, mondelli2019fundamental, luo2019optimal}. Then light is focused and TM fidelity is quantified based on the SBR of the generated foci. While in theory the transition is expected for $\alpha \simeq 4$ -- $5$, we see with an experimental dataset that $\text{SBR} > 100$ around $\alpha \simeq 10$ and even higher around $\alpha \simeq 15$. We also observe that the convergence is not the same for all the targets, maybe because all the targets are not equally excited due the Gaussian envelope of the illumination.

\subsection{TM Refinement via Light Refocusing}
Once the computational steps NMF + PR on experimental data $I_{out}$ are performed, a first reconstruction for $T$ is obtained.
\label{supp_varCor}
\hspace{5mm}
\begin{figure}[t!]
\centering
\includegraphics[width=\linewidth] {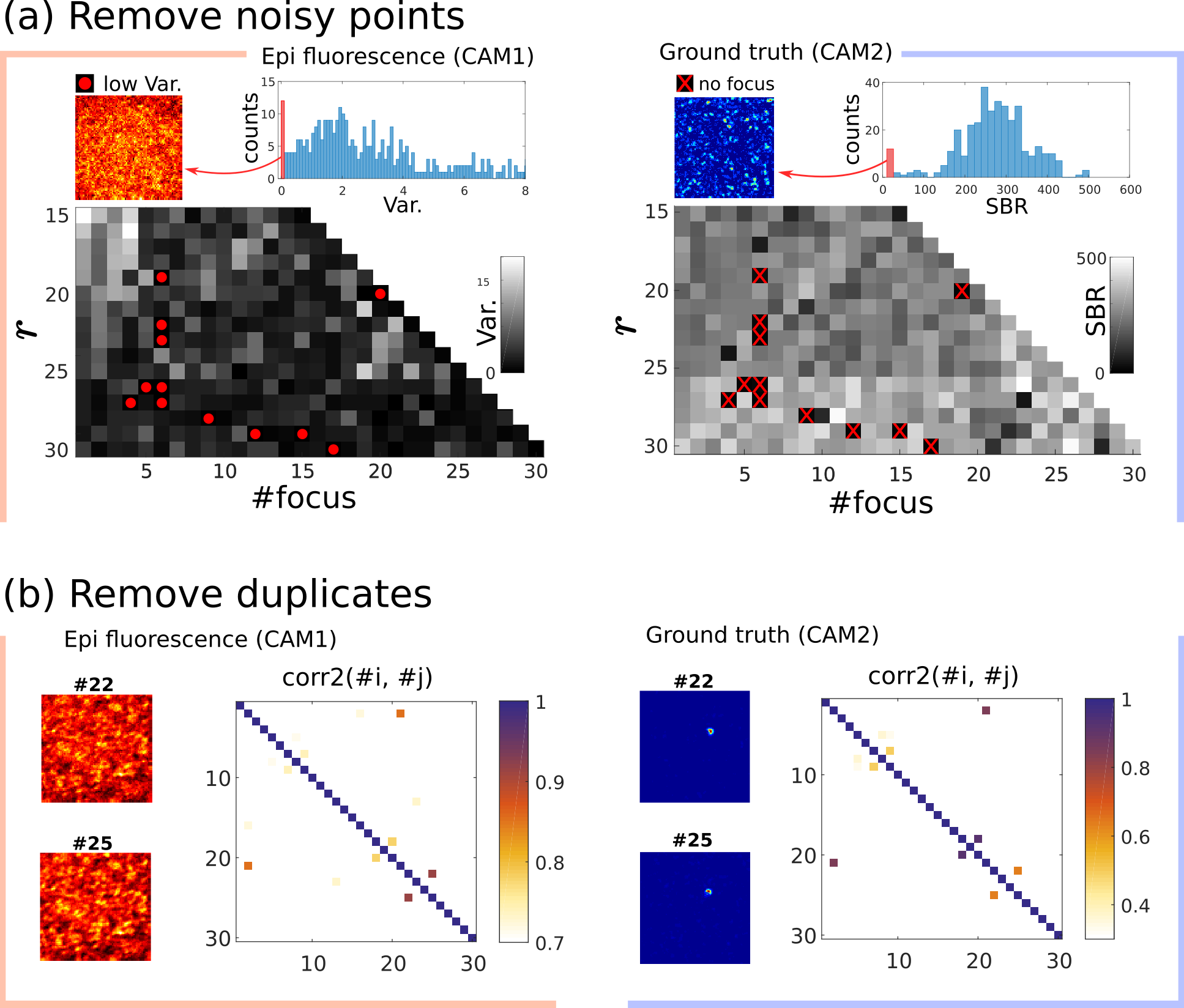}
    \caption{(a) Removing noisy points. When light is not successfully focused in transmission on one of the $N$ targets the corresponding fluorescence speckle pattern has a low spatial variance. (b) Removing duplicates. If two lines of the matrix $T$ focus on the same target the emitted fluorescence, is similarly back-scattered by the medium. The epi-detected speckles are thus strongly correlated.
    }
\label{Var_Corr}
\end{figure}
For several reasons mainly due to experimental noise, $T$ may not be perfectly reconstructed, with two consequences. First the reconstruction may fail: after taking the phase conjugation of those eigenvalues no focus is obtained at all. Second, the reconstruction may generate different eigenvalues focusing on the same target: these are duplicates. In the two cases, one needs to remove the corresponding line to improve the quality of $T$. Importantly this must be done non-invasively, by only looking at the epi-detected fluorescence.
In the first case, noisy points may be removed by looking at the spatial variance of the fluorescence speckle, as in \cite{Boniface:19}. In FIG. \ref{Var_Corr}a, we show that the fluorescence speckle having the lowest spatial variance corresponds to the lowest SBR points, i.e. speckle pattern instead of foci. To remove duplicates, we can perform the two-dimensional spatial correlation, FIG. \ref{Var_Corr}b. Here also, there is a good correspondence between what one can see in transmission on CAM2 and what can be epi-detected on CAM1.

\subsection{Memory Effect Characterization}
\label{subME}
\begin{figure}[t!]
\centering
\includegraphics[width=\linewidth] {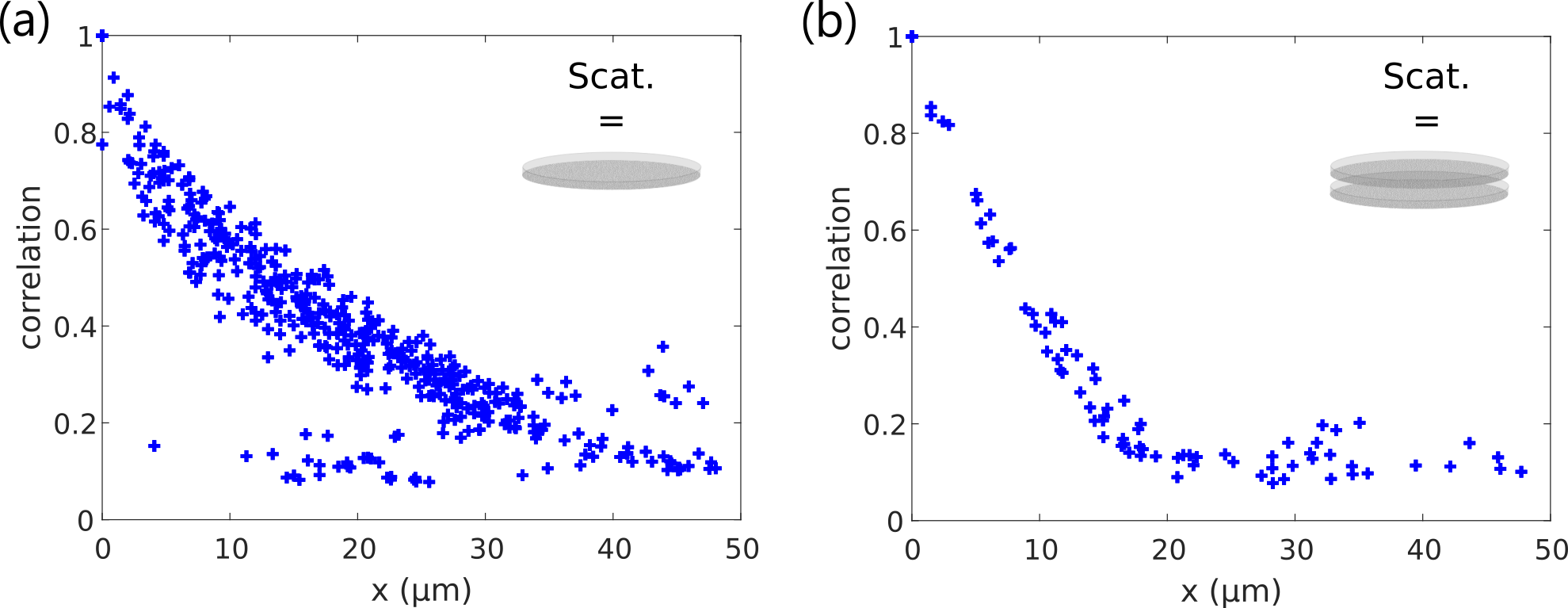}
    \caption{(a)-(b) Speckle spatial correlations due to ME inside the scattering medium respectively used for FIG. \ref{results} and FIG. \ref{imaging}.
    }
\label{MEsupp}
\end{figure}

\hspace{5mm}
The two experiments reported in FIG. \ref{results} and FIG. \ref{imaging} of the manuscript are conducted with two different scattering media. In the first case, our ability to consistently focus light in transmission is demonstrated through a single diffuser. From the fluorescence fingerprints we can perform some correlation measurements. With CAM2, placed in transmission we know the distance between the sources emitting the former patterns. Altogether, the two information can be used to estimate the memory effect range for the linear fluorescence at 540/560nm. Memory effect for the excitation light at 532nm is expected to be similar. Graph is reported on FIG. \ref{MEsupp}a. From the graph, we can roughly estimate that memory effect range does not exceed 40\textmu m. The object has a similar size and such memory effect range is enough to recover the full object in one shot using an autocorrelation technique, \cite{katz2014non}. 
In order to show that our non-invasive imaging technique, is not limited to a single memory effect patch, we decided to work in a stronger scattering regime by using two holographic diffusers separated by $0.78$mm (the thickness of the diffuser). In this situation, the memory effect range is reduced to $\simeq$20\textmu m, FIG. \ref{MEsupp}b. We use an extended object, whose typical size is $\simeq$50\textmu m. In this configuration the autocorrelation technique cannot be applied.  

\subsection{PR Necessity for Imaging}
\label{supp_dist}
\hspace{5mm}
To reconstruct the object, as done in FIG. \ref{imaging}, we exploit the spatial correlations between the fluorescent speckles. In principle, the latter can be retrieved via two different methods: NMF or NMF + PR. In the following, we discuss the differences.

\begin{figure}[htbp]
\centering
\includegraphics[width=\linewidth] {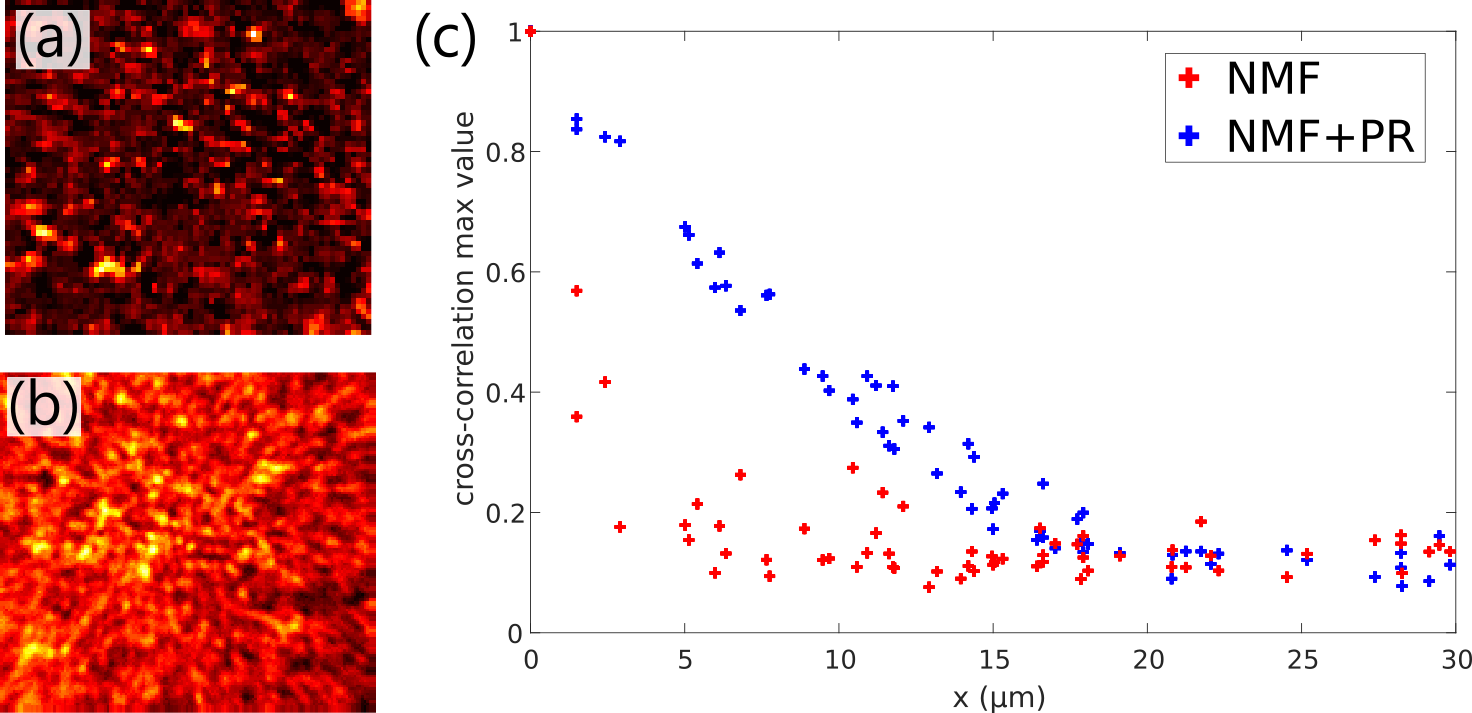}
    \caption{(a) Example of pattern obtained via the NMF. (b) Epi-detected fluorescence patterns emitted by the beads when light is focused, NMF + PR. (c) Correlation between patterns (obtained via NMF and NMF + PR) as a function of the distance between the fluorescent emitters.}
\label{NMFdistance}
\end{figure}

On one hand, the NMF should directly provide the fluorescent eigen-patterns corresponding to each bead, with factor matrix $W$.
However before running the NMF, a Gaussian filter is used to remove structures with spatial frequencies lower than the speckle grain size. The idea is to remove both the detection noise and the fluorescence background envelope. As also discussed in \cite{moretti2019readout}, this step seems to be crucial to obtain a reliable demixing through the NMF. As a consequence, the computationally obtained patterns (FIG. \ref{NMFdistance}a) do not contain all the information of fluorescence back-scattering, but only a filtered version. Whereas it is not a problem to retrieve the TMs, it severely impacts the reconstruction of the object.

On the other hand, by reconstructing the matrix $T$ through the NMF + PR pipeline, selective focusing on all the beads can be done thanks to phase conjugation of the TM. While light is focused on the beads, their fluorescence patterns can be non-invasively recorded (FIG. \ref{NMFdistance}b).

Finally, cross-correlations between patterns obtained through the same method (NMF or NMF + PR) are performed. In the case of NMF, part of the information is missing which explains why patterns have some correlation only if the sources are very close ($<5$\textmu m), see FIG. \ref{NMFdistance}c. In the other case, NMF + PR, we directly use the raw patterns measured on the camera, and correlation is maintained over a much larger region ($\sim 15$\textmu m). Note that the latter is limited by the memory effect of the scattering medium.

\subsection{MDS Reconstruction}
\label{subMDS}
\begin{figure}[t!]
\centering
\includegraphics[width=\linewidth] {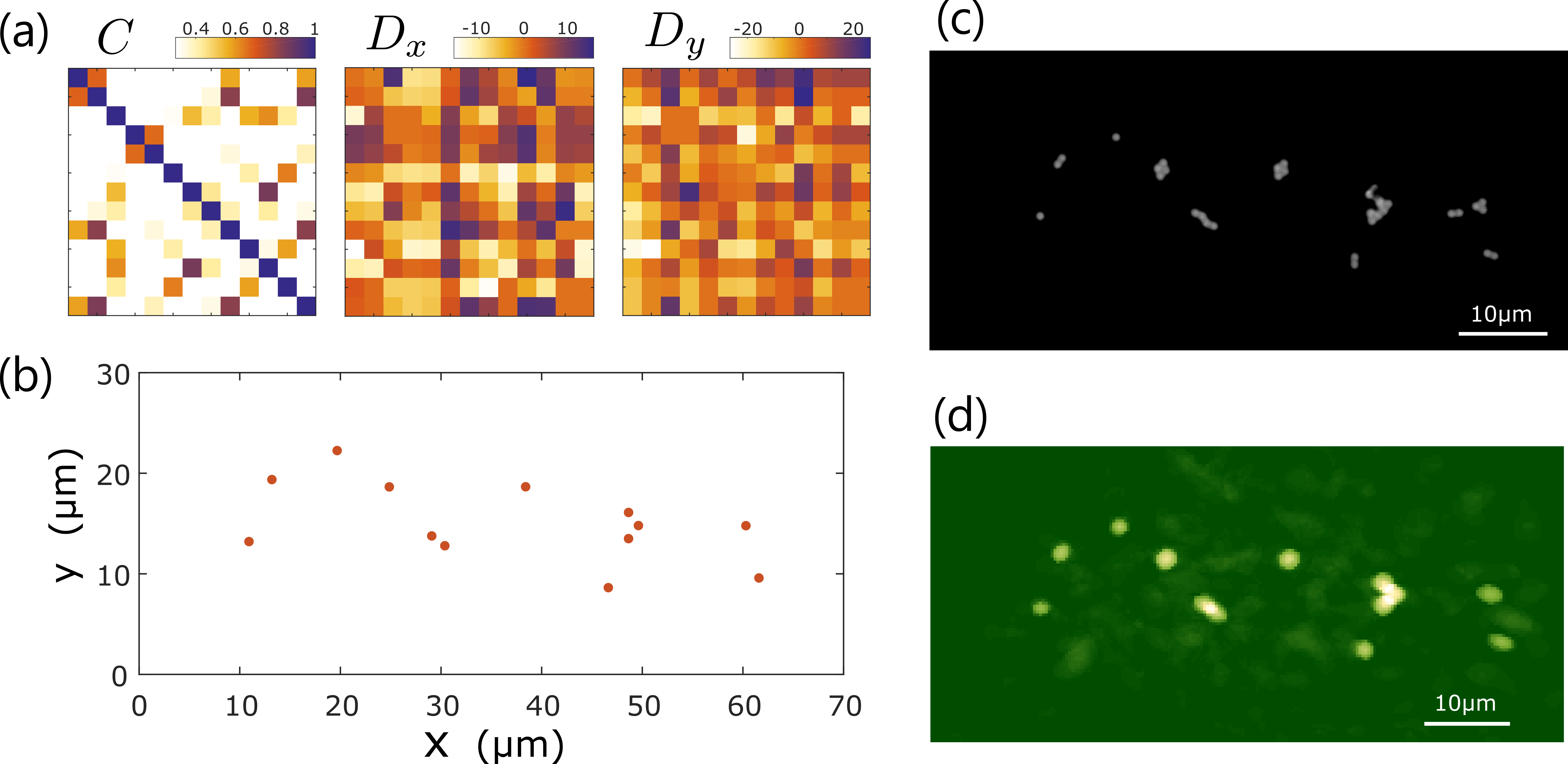}
    \caption{(a) Estimated matrices $C$, $D_x$ and $D_y$, from fluorescent speckle cross-correlations. (b) MDS reconstruction. (c) Ground truth fluorescence image obtained without scattering medium. (d) Patch by patch reconstruction.}
\label{MDSsupp}
\end{figure}

\hspace{5mm}
As described in FIG. \ref{imaging}, cross-correlation gives access to the translation $\overrightarrow{u_{ij}}$ which corresponds to the relative shifts along the x and y-axis between beads $i$ and $j$. 
Once all the pairwise cross-correlations $\{i,j\}$ are calculated, we end up with three matrices:
\begin{itemize}
    \item $C$, the correlation matrix which corresponds to the intensity of the off-centered cross correlation peak.
    \item $D_x$, the horizontal distance matrix, which stands for the estimated x-axis shift between all the beads.
    \item $D_y$, the vertical distance matrix, which stands for the estimated y-axis shift between all the beads.
\end{itemize}
From there, different strategies for the reconstruction may be considered. The one presented in the manuscript consists in doing the reconstruction patch by patch. One major limitation is that to retrieve the position of one point only the its close neighbour (with strong correlation) are used. Herewith we propose a global approach relying on Multi-Dimensional Scaling (MDS) \cite{cox2008multidimensional}. In this case, the full distance matrix is used through an optimization algorithm which is more robust to noise than the approach proposed in the manuscript. The stress function is weighted with $c_{ij}^{\alpha}$, where $c_{ij}$ are the correlation matrix coefficients and reads:
\begin{equation*}
    stress_{[u_1,u_2,...,u_r]}=\left(\sum_{i\ne j=1...r}{c_{ij}^{\alpha}(d_{ij}-\norm{u_i-u_j})^2}\right)^{1/2}
\end{equation*}
The optimization is actually run twice, to successively access all the $u_i=x$ and $u_i=y$ coordinates for $i = 1, \ldots,r$, with distance matrix $D_x$ and $D_y$ respectively.
Retrieved positions are plotted in FIG. \ref{MDSsupp}a for $\alpha=6$. This new approach is also in good agreement with the ground truth (b). 
Here, in the case of a relatively simple object, the MDS method provides a reconstruction very similar to the patch by patch approach (c). The two images are obtained from the same dataset but through a different reconstruction algorithm.

\subsection{Numerical Results with 2-photon Excitation}
\hspace{5mm}
A major advantage of our approach is that it should be applicable to any incoherent process. In the following, we provide some simulation for 2-photon fluorescence.
\label{sub2p_cont}
\begin{figure}[htbp]
\centering
\includegraphics[width=\linewidth] {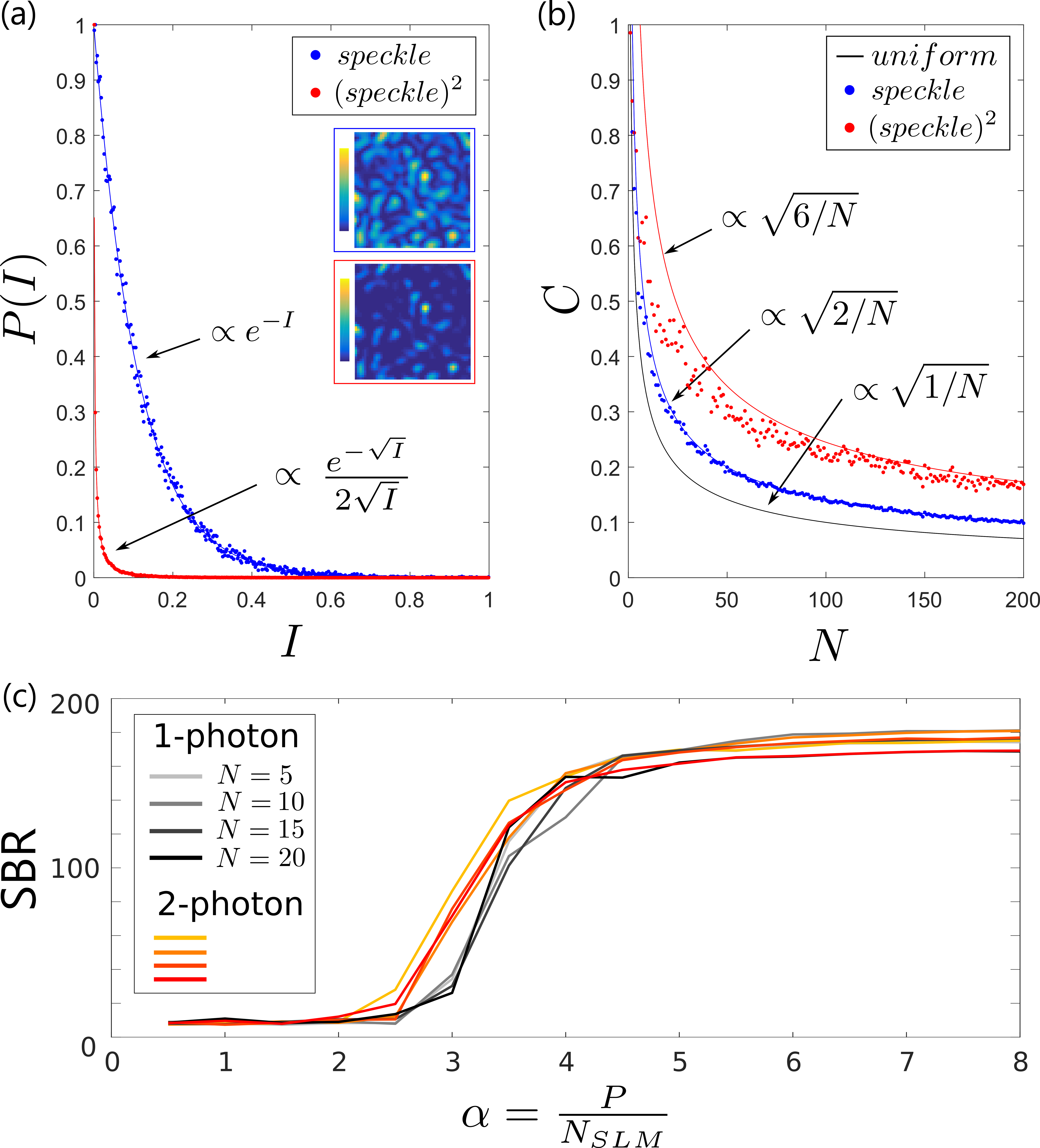}
    \caption{(a) Probability density function respectively derived for, a standard $speckle$ defined by intensity $I$, and a 2-photon speckle, denoted $(speckle)^2$ obtained by squaring the intensity of a standard $speckle$.
    (b) Contrast evolution of fluorescence speckle emitted by $N$ targets. The $N$ targets are excited with different distribution: $uniform$, $speckle$ and $(speckle)^2$. (c) NMF + PR algorithm is used in the two cases to retrieve matrix $T$ and focus light with a different number of inputs. Whatever the number of targets $N$ the 2-photon configuration proves to be faster in terms of convergence.
    }
\label{2Psupp}
\end{figure}
This additional non-linearity actually helps our algorithm, since spatial sparsity is even higher. As one can see on FIG. \ref{2Psupp}a, if speckle intensity is squared, the spatial sparsity is significantly increased. The corresponding speckle is composed by more dark speckle grains and the probability density function is more strongly peaked around zero intensity. The probability to get only few targets excited at once is increased which should make the fluorescence demixing easier. Additionally, such an illumination pattern also improved the contrast of the back-scattered fluorescent. Theory predicts that the contrast scales as $\propto \sqrt{6/N}$ instead of $\propto \sqrt{2/N}$ for 1-photon fluorescence (see after for detailed demonstration), FIG. \ref{2Psupp}b. It would give the opportunity to increase the number of targets with a similar contrast. These two effects actually help the algorithm to converge. We report on FIG. \ref{2Psupp}c, simulations for the two different regimes. No noise is added and $N_{SLM}=256$. 

\hspace{5mm} 
We now derive the contrast of the sum of $N$ non equal strength speckle intensities. We assume that the $N$ individual speckles are statistically independent.

The total intensity of interest is given by the sum
$ I_s=\sum_{n=1}^{N}{I_n} $
where $ I_n $ has a mean value $ \bar{I_n} $. We consider here that each corresponding speckle is fully developed and obeys negative exponential probability distribution. In particular it results in the fact that its mean intensity is equal to its standard deviation, $ \sigma_n = \bar{I_n}  $. 
The mean value of the total intensity thus reads
$ \bar{I_s} =\sum_{n=1}^{N}{\bar{I_n}} $.

Then, the second moment of the total intensity is
$ \sigma_s^2 = \sum_{n=1}^{N}{\sigma_n^2}  $
and the contrast of the total intensity is
\begin{equation*}
C = \frac{\sigma_s}{\bar{I_s}} = \frac{\sqrt{\sum_{n=1}^{N}{\sigma_n^2}}}{\sum_{n=1}^{N}{\bar{I_n}}}
\end{equation*}

If all the $N$ components have an equal mean intensity ($ \bar{I_n} = I_0 $), the previous equation reduces to the well-known expression 
$ C = \frac{1}{\sqrt{N}}$.

Now we look into the case where the mean intensity $ \bar{I_n} $ is not the same for all the $N$ components but rather follows a negative exponential probability distribution. Such speckles are obtained when spatially incoherent beads are excited with a speckle illumination.
In this situation the contrast follows
$ C = \frac{1}{\sqrt{N}} \frac{\sqrt{1/N\sum_{n=1}^{N}{\sigma_n^2}}}{1/N\sum_{n=1}^{N}{\bar{I_n}}}$ 
and tends to 
$ C = \frac{1}{\sqrt{N}} \frac{\sqrt{\int_{0}^{\infty}{I^2 e^{-I} dI}}}{\int_{0}^{\infty}{I e^{-I} dI}} = \sqrt{\frac{2}{N}} $ 
when N tends to infinity.

\newpage
\renewcommand{\refname}{References} 
\bibliography{BIBLIO}

\end{document}